\documentclass[conference]{IEEEtran}
\usepackage{algorithm,amsbsy,amsmath,amssymb,epsfig,bbm,mathrsfs,fancyhdr,fancyvrb,url,color,soul}
\usepackage{amsthm}
\usepackage{graphicx}
\usepackage{cite}
\usepackage{float}
\usepackage{mathtools}
\usepackage{algorithm}
\usepackage{algpseudocode}
\usepackage{epstopdf}
\usepackage{epsfig}
\usepackage{tabularx}
\usepackage[caption=false,font=footnotesize]{subfig}
\usepackage{bm}
\usepackage{multicol}
\usepackage{makecell}
\usepackage{afterpage}
\usepackage{dsfont}

\usepackage{caption}
\usepackage[font=small]{caption} 
\usepackage{array}

\usepackage{xcolor}

\usepackage[english]{babel}

\newcolumntype{M}[1]{>{\centering\arraybackslash}m{#1}}

\IEEEoverridecommandlockouts
\title{RSMA-Assisted OFDM–OTFS Hybrid Framework for Mixed-Mobility Multiuser Systems}
\author{\IEEEauthorblockN{Wafa Hedhly, Leila Musavian, and Nikolaos Thomos}
 \thanks{The authors are with the School of Computer Science and Electronic Engineering, University of Essex, Wivenhoe Park, Colchester CO4 3SQ, United Kingdom (e-mail: { \{wafa.hedhly, leila.musavian, nthomos\}@essex.ac.uk).} This work was funded by the Engineering and Physical Science Research Council, UK, [grants number EP/X012204/1, EP/X04047X/2, EP/Y037243/1].}
 }
\usepackage{stfloats}
\usepackage{amsmath}

\begin{document}
\maketitle

\begin{abstract}
   In future 6G vehicular networks, users employing orthogonal frequency division multiplexing (OFDM) and orthogonal time frequency space (OTFS) waveforms may coexist under diverse mobility conditions, where both can experience high-mobility and low-mobility profiles. Since OFDM users can suffer severe inter-carrier interference (ICI) and OTFS users occupy larger spectrum resources, rate-splitting multiple access (RSMA) is a flexible framework that can efficiently handle these heterogeneous aspects.
   In this work, we propose a novel RSMA-assisted system to provide downlink communication to multiple OFDM and OTFS users. A common stream comprising the common messages of OFDM users spans the whole bandwidth to help OFDM users manage the ICI induced by potential high Doppler effects. OTFS users do not participate in the common stream. 
   The private streams of OFDM users and the streams of OTFS users are transmitted over disjoint frequency bands.  
    During the SIC process implemented at all receivers, channel estimation errors are taken into account. 
    The simulation results highlight the impact of the power allocation factors and channel estimation errors on the system performance, and demonstrate the superiority of the proposed framework over orthogonal multiplexing in terms of outage probability and rate performance.
\end{abstract}

\section{Introduction}
 Future vehicular networks are expected to handle massive numbers of connected devices, moving at different speeds and with diverse quality-of-service (QoS) requirements. Therefore, suitable waveforms and efficient interference management and spectrum utilization play an important role in order to ensure reliable and sustainable networks \cite{chen2011fundamental}. 
 Although orthogonal frequency division multiplexing (OFDM) has been widely adopted for its compatibility with existing infrastructures, its sensitivity to Doppler effects leads to significant performance degradation in high-mobility scenarios \cite{park2023intercarrier}. Orthogonal time frequency space (OTFS) has emerged as a promising waveform capable of operating effectively in double-dispersive channels \cite{raviteja2018interference,ding2019otfs,hedhly2024otfs}. Nevertheless, due to the undeniable merits of OFDM, OTFS has not yet been fully adopted leading to the inevitable coexistence of OFDM-legacy devices and OTFS devices in future networks. Therefore, multiple access techniques are expected to efficiently enable the coexistence of users operating in the time-frequency (TF) domain and others operating in the delay-Doppler (DD) domain. 

Rate-splitting multiple access (RSMA) has been considered a middle ground between fully orthogonal and fully non-orthogonal access techniques \cite{mao2022rate,clerckx2016rate}. RSMA exhibits better interference management than non-orthogonal multiple access (NOMA) by partially decoding and treating interference as noise, as well as lower receiver complexity due to fewer stages of successive interference cancellation (SIC) \cite{mao2022rate}. 
RSMA has been considered a suitable multiple access technique to be coupled with OTFS in order to serve users with different mobility profiles \cite{huai2024cross,liu2024power,huai2025downlink,costa2025refined}. A power allocation scheme was developed in \cite{liu2024power} to improve spectral efficiency in a high-mobility OTFS-RSMA system with a path-selective one-tap receiver, where all users' signals are represented in the DD domain. 
 The work in \cite{huai2024cross} proposed an uplink OTFS–RSMA cross-domain transmission scheme to serve a high-mobility user and a low-mobility user, enhancing the performance in terms of outage and fairness.
 The work in \cite{huai2025downlink} considered a downlink scenario, optimizing rate and power allocation to maximize the sum-rate. In the proposed architecture, the high-mobility users share a common stream represented in the DD domain, while low-mobility users share a distinct common stream represented in the TF domain. 
 The authors of \cite{costa2025refined} proposed a fairness-aware resource allocation scheme in OTFS–RSMA in LEO ISAC systems, under imperfect channel state information (CSI) conditions. The downlink users are all moving at high speeds and modulated using OTFS.
 In contrast, the aforementioned research works considering OTFS-RSMA are limited to all high-mobility users assumptions or domain-separation of common streams. A general framework for the scenario of mixed users mobility profiles with mixed waveforms is needed.
Another group of works studied the integration of RSMA with OFDM systems to improve interference management under intercarrier interference (ICI) \cite{csahin2023multicarrier,csahin2024ofdm,rajkumar2025power}. In \cite{csahin2023multicarrier,csahin2024ofdm}, the authors proposed OFDM-RSMA frameworks under doubly selective channels, where power and subcarrier allocation were optimized to manage ICI. The work in \cite{rajkumar2025power} investigated RSMA-OFDM with amplify-and forward (AF) relaying considering practical impairments such as phase noise and ICI. Although these studies considered OFDM users operating in Doppler channels, they did not investigate the coexistence with other waveforms.

In mixed-mobility networks, the base station (BS) is expected to serve users supporting both OFDM and OTFS waveforms and moving at different speeds. 
Thus, RSMA is  particularly attractive in mixed OFDM-OTFS scenarios.
On one hand, the enhanced interference management of RSMA helps improve the system performance where high-mobility OFDM users may experience significant inter-carrier interference. On the other hand, the reduced complexity of RSMA SIC process facilitates the co-existence of OTFS and OFDM receivers by reducing the complexity of equalization in different domains.
In this paper, we propose an RSMA-assisted framework to serve users operating with OFDM-legacy receivers and OTFS receivers in an interference-limited environment. In the considered downlink system, OFDM users and OTFS users simultaneously share the available spectrum under heterogeneous channel conditions. In order to manage ICI, OFDM users receive parts of their messages as a common stream spanning the entire bandwidth, while their private streams are allocated to disjoint subcarriers. OTFS users occupy separate frequency bands and operate in the DD domain. OTFS users cancel the common stream before decoding their own streams. Moreover, the impact of ICI on OFDM transmissions and imperfect channel estimation errors during SIC are explicitly considered in the system model. Numerical results demonstrate that the proposed RSMA framework improves the total sum-rate performance. It also improves the rate and outage performance of OFDM users while maintaining flexible coexistence between OFDM and OTFS transmissions. The results also reveal an inherent trade-off between OFDM reliability and OTFS throughput depending on the common-stream power allocation.

\section{System Model}
In this work, we consider a downlink (DL) communication where a transmitter (Tx) is communicating with $K_1$ OFDM users and 
$K_2$ OTFS users. The Tx and all users are equipped with a single antenna. In this setup, OTFS users are served in the DD domain and OFDM users are served in the TF domain. We consider $M$ available subcarriers.
The Tx accommodates OFDM users using the RSMA scheme and hence, splits their signals into a common stream and private streams. Each private stream occupies a single subcarrier, and the common stream is transmitted over the whole spectrum. 
Signals transmitted to OTFS users are separately encoded in OTFS frames and transmitted over the remaining $M-K_1$ subcarriers. Therefore, OTFS users suffer interference from the common stream and remove it through SIC. We assume that each OTFS user occupies $L$ subcarriers. We denote by $\mathcal{M}_u$ the set disjoint subcarriers occupied by user $u$, $0 \le u \le K_1+K_2-1$.

We denote by $W_u$ the message of user $u, 0 \le u \le K_1 + K_2 -1 $. The message $W_u$ of OFDM user $u$ is split at the transmitter in a common message $W_{\mathrm{c},u}$ and a private message $W_{\mathrm{p},u}$. The Tx combines the common messages of all OFDM users into one common stream $W_\mathrm{c}$. The private messages and common messages are all separately encoded. We denote by $s_\mathrm{c}$ the encoded common stream, $s_{u}, K_1 \le u \le K_1+K_2 -1$ the OTFS user encoded stream and $s_u, 0 \le u \le K_1-1$ the encoded private streams of OFDM user $u$. 
The Tx allocates a common rate portion $C_u$ to user $u$ in the common stream.  
The Tx total transmit power is $P_\mathrm{t}$. The common-stream power is equally distributed over all subcarriers, i.e., $p_{\mathrm{c},m} = p_{\mathrm{c}0}$, $\forall m$. 
 The parameter $p_u$ denotes the power allocation factor of user $u$, where $0 \le u \le K_1+K_2 - 1 $. For OFDM users, $p_u$ corresponds to the power assigned to their private streams.


The transmitted signal over subcarrier $0 \le m \le M-1$ can be expressed as,
\begin{equation}
    s(m) = \sqrt{p_{\mathrm{c}0}} s_\mathrm{c}(m) +  \sum_{u = 0}^{K_1 + K_2 - 1} \sqrt{p_u} s_u(m),
\end{equation}
where,
\begin{align}
s_u(m) =
\begin{cases}
s_u, & m = u,\; 0 \le u \le K_1-1, \\
s_u(m), & u \ge K_1,\; m \in \mathcal{M}_u.
\end{cases}
\end{align}

The Tx transmits an $N \times L$ OTFS frame composed of $NL$ information symbols to the OTFS user, where $N$ is the number of Doppler bins and $L$ is the number of delay bins. If $T$ is the symbol duration and $\Delta f = 1/T$ the subcarrier spacing, the resulting packet has a total duration of $NT$ and a bandwidth of $L \Delta f$. 
The TF domain representation of the OTFS signal $x_u^{\mathrm{TF}}$ at instant $n$ and subcarrier $m$ can be obtained by applying the ISFFT to its DD-domain representation $x_u^{\mathrm{DD}}$ as follows,
\begin{equation}
    x_u^{\mathrm{TF}}[n,m] = \frac{1}{\sqrt{NL}} \sum_{k = 0}^{N-1} \sum_{l = 0}^{L-1} x_u^{\mathrm{DD}}[k,l] e^{j2\pi \left( \frac{kn}{N}-\frac{ml}{L}  \right)},
\end{equation}
where $0 \le k \le N-1$, $0 \le l \le L-1$.

We denote by $h_{p,u}$, $\tau_{p,u}$ and $\nu_{p,u}$ the channel gain, delay and Doppler of path $p$. $1 \le p \le P_u$, where $P_u$ is the number of channel paths of user $u$.
In the DD domain, for an OTFS user, each path $p$ is represented by a pair of delay tap $l_{p,u}$ and Doppler tap $k_{p,u}$. The Doppler and delay of path $p$ are therefore expressed as,
\begin{equation}\label{taps}
    \nu_{p,u} = \frac{k_{p,u}}{NT} \quad \mathrm{and} \quad \tau_{p,u} = \frac{l_{p,u}}{L \Delta f}.
\end{equation}
The path gains $h_{p,u}, 1 \le p \le P_u $ are independent and identically distributed (i.i.d) $\mathcal{C}(0,1/P_u)$. 

The channel impulse response in the TF domain when Doppler effects are considered is obtained as,
\begin{equation}
\mathbf{H}_u^{\mathrm{TF}} = \mathbf{F}_M \mathbf{R}_{\mathrm{CP}} \left( 
\sum_{p=1}^{P_u}
h_{p,u} \mathbf{\Delta}({\nu_{p,u}}) \mathbf{\Pi}^{\tau_{p,u} F_s} \right) \mathbf{A}_{\mathrm{CP}} \mathbf{F}_M^\mathrm{H},
\end{equation}
where $F_\mathrm{s} = M \Delta f$ is the sampling frequency, $\mathbf{F}_M$ is the discrete Fourier transform (DFT) matrix, $\mathbf{R}_{\mathrm{CP}}$ is the cyclic prefix (CP) removal matrix and $\mathbf{A}_{\mathrm{CP}}$ is the CP insertion matrix obtained as,
\begin{align}
\mathbf{A}_{\mathrm{CP}} &=
\begin{bmatrix}
\mathbf{0}_{N_{\mathrm{CP}}\times(M-N_{\mathrm{CP}})} & \mathbf{I}_{N_{\mathrm{CP}}} \\
\mathbf{I}_{M}
\end{bmatrix}, \\
\mathbf{R}_{\mathrm{CP}} &=
\begin{bmatrix}
\mathbf{0}_{M\times N_{\mathrm{CP}}} & \mathbf{I}_{M}
\end{bmatrix},
\end{align}
where $N_{\mathrm{CP}}$ is the CP length of each user chosen according to the delay spread of the channel. The matrix $\mathbf{\Pi}$ represents a circular delay shift, while $\mathbf{\Delta}$ is a diagonal Doppler-shift matrix obtained as,
\begin{equation}
\mathbf{\Delta}(\nu_{p,u}) =
\mathrm{diag}
\left(
1,
e^{j2\pi \frac{\nu_{p,u}}{F_s}},
e^{j2\pi \frac{2\nu_{p,u}}{F_s}},
\ldots,
e^{j2\pi \frac{(M-1)\nu_{p,u}}{F_s}}
\right)
\end{equation}
We consider the following notation $h_u(m,j) = [H_u]_{m,j}^{\mathrm{TF}}$.

\section{Signals Detection for the Proposed RSMA-OTFS Paradigm}
\subsection{Received Signals at OFDM Users}

Since the SIC depends on the channel estimation, it is important to investigate the effect of channel estimation errors on the residual interference and on the system performance in general. We assume that over each subcarrier $m$ and for every OFDM user $u, 0 \le u \le K_1-1$, the CSI estimation process incurs an error modeled as follows,
\begin{equation}
    {h}_{u}(m,j) = {\hat{h}}_u(m,j) + {e}_u(m,j), \quad 0 \le u \le K_1-1,
\end{equation}
where ${\hat{h}}_u(m,j) \sim \mathcal{CN}(0, \sigma_{h_{u,m}}^2)$ is the estimated channel coefficient and ${e}_u(m,j) \sim \mathcal{CN}({0}, \sigma_{e_u}^2)$  denotes the channel error \cite{li2019residual}. The term $\hat{h}_u(m,m)$ is the desired channel and $\hat{h}_u(m,j), j \neq m$ is the ICI from subcarrier $j$ to subcarrier $m$. We assume that the delay and Doppler of each path are known at the receiver and estimation errors mainly affect the path gains.

The received signal at user $u, 0 \le u \le K_1-1$ over subcarrier $m$ can be expressed as,                        
\begin{equation}
\begin{aligned}
y_{\mathrm{c},u}(m)
&= \sqrt{p_{\mathrm{c}0}}\, h_u(m,m)s_{\mathrm{c}}(m)+ \sum_{\substack{j=0\\j\neq m}}^{M-1}
\sqrt{p_{\mathrm{c}0}}\, h_u(m,j)s_{\mathrm{c}}(j) \\
&\quad + \sum_{v=0}^{K_1+K_2-1}
\sum_{j\in\mathcal{M}_v}
\sqrt{p_v}\, h_u(m,j)s_v(j) + n_u(m).
\end{aligned}
\end{equation}
where $n_u(m)$ is the additive Gaussian noise over subcarrier $m$ with variance $\sigma^2$.

The instantaneous SINR for decoding the common stream at user $0 \le u \le K_1-1$ over subcarrier $m$ is written as,
\begin{equation}
\gamma_{\mathrm{c},u}(m)=
\frac{
\gamma_{\mathrm{t}}p_{\mathrm{c}0}|h_u(m,m)|^2
}{
\gamma_{\mathrm{t}}  I_{\mathrm{c},u}(m)+1
},
\end{equation}
where $\gamma_{\mathrm{t}} = P_{\mathrm{t}}/\sigma^2$ and $I_{\mathrm{c},u}$ is the overall interference term expressed as follows,
\begin{equation}
I_{\mathrm{c},u}(m)
=
p_{\mathrm{c}0}
\sum_{\substack{j=0\\j\neq m}}^{M-1}
|h_u(m,j)|^2
+
\sum_{v=0}^{K_1+K_2-1} p_v
\sum_{j\in \mathcal{M}_v}
 |h_u(m,j)|^2 .
\end{equation}
The term $I_{\mathrm{c},u}(m)$ includes the same-subcarrier data interference from the signal transmitted over subcarrier $m$, as well as the ICI leakage from all other subcarriers, including both the common stream and the users' data streams.
After applying the SIC at each OFDM user $u$, for $0 \le u \le K_1-1$, the received signal over the dedicated subcarrier $m=u$ is expressed as follows,
\begin{equation}
\begin{aligned}
y_{u,\mathrm{r}}(u) &=
\sum_{j=0}^{M-1} \sqrt{p_{\mathrm{c}0}}\, e_u(u,j)\, s_\mathrm{c}(j) + \sqrt{p_u}\, h_u(u,u)\, s_u(u) \\
&\quad + \sum_{\substack{v=0\\v\neq u}}^{K_1+K_2-1}
\sum_{j\in\mathcal{M}_v}
\sqrt{p_v}\, h_u(u,j)\, s_v(j)  + n_u(u).
\end{aligned}
\end{equation}
where $\sqrt{p_{\mathrm{c}0}} e_u(u,j) s_\mathrm{c}(j)$ is the residual signal from the imperfect SIC from each subcarrier $j$. 

Thus, the instantaneous SNR for decoding the private stream at OFDM user $u$ where $0 \le u \le K_1-1$ is expressed as,
\begin{equation}
\gamma_{\mathrm{p},u}(u)=
\frac{
\gamma_{\mathrm{t}}p_u|h_u(u,u)|^2
}{
\gamma_{\mathrm{t}}p_{\mathrm{c}0}
\sum_{j=0}^{M-1}|e_u(u,j)|^2
+
\gamma_{\mathrm{t}} I_{\mathrm{p},u}(u)
+
1
}.
\end{equation}
where $I_u$ is the ICI term expressed as follows,
\begin{equation}
I_{\mathrm{p},u}(u)=
\sum_{\substack{v=0\\v\neq u}}^{K_1+K_2-1}
 p_v\sum_{j\in\mathcal{M}_v}
|h_u(u,j)|^2 .
\end{equation}
\subsection{Received Signals at OTFS Users}
OTFS users first decode the common stream prior to detecting their private DD-domain signals.
According to the received signal in the DD-domain in \cite{hedhly2024otfs} and since each path gain can be expressed as a sum of estimated term and error term, the DD-domain channel matrix can be obtained as,
\begin{equation}
    \mathbf{H}_u = \hat{\mathbf{H}}_u + \mathbf{E}_u,
\end{equation}
where $\hat{\mathbf{H}}_u$ and $\mathbf{E}_u$ are $NL \times NL$ matrices.
After performing SIC to the common stream, the OTFS user $u$ observes the following signal in the DD domain,
 \begin{align}\label{y0}
    \mathbf{y}_u = \sqrt{p_u} \mathbf{H}_u \mathbf{s}_u + \sqrt{p_{\mathrm{c}0}} \mathbf{E}_u \mathbf{s}_\mathrm{c} + \mathbf{n}_u,
 \end{align}
 where $\mathbf{s}_u$ and $\mathbf{s}_\mathrm{c}$ denote the DD-domain vector representation of the transmitted signals over the $N \times L$ resource block and $\mathbf{n}_u$ is the DD-domain representation of the additive noise.

We apply the minimum mean squared error (MMSE) equalization to detect the OTFS signal in the DD domain. 
We start by denoting $\mathbf{F}_N$ and $\mathbf{F}_L$ as $N \times N$ and $L \times L$ unitary DFT matrices, respectively. We consider the following unitary matrix $\mathbf{\Psi} = \mathbf{F}_N \otimes \mathbf{F}_L$.

Thanks to the properties of block-circulant matrices, the matrices $\mathbf{H}_{u}$ and $\mathbf{E}_{u}$ can be diagonalized as follows, 
\begin{align}
    \mathbf{H}_{u} = \mathbf{\Psi}^\mathrm{H} \mathbf{\Delta}_u \mathbf{\Psi}, 
      \mathbf{E}_{u} = \mathbf{\Psi}^\mathrm{H} \mathbf{\Delta}_{\mathrm{e},u} \mathbf{\Psi},
\end{align}
where $\mathbf{\Delta}_u = \mathrm{diag} \left[ \delta_{1}, \ldots, \delta_{NL} \right]$ and $\mathbf{\Delta}_{\mathrm{e},u} = \mathrm{diag} \left[ \epsilon_{1}, \ldots, \epsilon_{NL} \right]$. 
Therefore, we apply MMSE equalizer obtained as,
\begin{equation}
\mathbf G_u = \mathbf H_u^H \left( \mathbf H_u\mathbf H_u^H + \rho \mathbf I_{NL} \right)^{-1}, \qquad \rho=\frac{1}{\gamma_t}.
\end{equation}
Thus, the equalized signal in the DD domain can be expressed as,
\begin{align}
\hat{\mathbf y}_u &=
\sqrt{p_u}\, \boldsymbol{\Psi}^{H}
\mathbf G_{\Delta,u} \boldsymbol{\Delta}_u
\boldsymbol{\Psi} \mathbf s_u + \sqrt{p_c}\, \boldsymbol{\Psi}^{H}
\mathbf G_{\Delta,u} \boldsymbol{\Delta}_{e,u} \boldsymbol{\Psi}
\mathbf s_c \nonumber \\
&+ \boldsymbol{\Psi}^{H}
\mathbf G_{\Delta,u}
\boldsymbol{\Psi}
\mathbf n_u ,
\end{align}
where $G_{\Delta,u} = \mathrm{diag} \left[\frac{\delta_{u,1}^*}{ |\delta_{u,1}|^2 + \rho} , \ldots, \frac{\delta_{u,NL}^*}{ |\delta_{u,NL}|^2 + \rho}  \right].$ 

Thus, the SINR for detecting the OTFS signal of user $u$ in the DD domain is obtained as,
\begin{equation}
    \gamma_u = \frac{\gamma_t p_u \alpha_u^2}{\gamma_t p_u \beta_u+\gamma_t p_c \omega_{e,u}+\omega_{0,u}},
\end{equation}
where,
\begin{align}
\alpha_u =\frac{1}{NL}\sum_{i=1}^{NL}\frac{|\delta_{u,i}|^2}{|\delta_{u,i}|^2+\rho}, 
\beta_u=\frac{1}{NL}\sum_{i=1}^{NL}\left|\frac{|\delta_{u,i}|^2}{|\delta_{u,i}|^2+\rho}-1 \right|^2,
\end{align}
\begin{align}
\omega_{0,u}=\frac{1}{NL}\sum_{i=1}^{NL}\frac{|\delta_{u,i}|^2}{\left(|\delta_{u,i}|^2+\rho\right)^2},
\omega_{e,u}=\frac{1}{NL}\sum_{i=1}^{NL}\frac{|\delta_{u,i}|^2|\epsilon_{u,i}|^2}{\left(|\delta_{u,i}|^2+\rho\right)^2}. 
\end{align}

The instantaneous rate for decoding the common stream at OFDM user $u$, where $0 \le u \le K_1-1$ is expressed as follows,
\begin{equation}
    R_{\mathrm{c},u} = \sum_{m = 0}^{M-1} \log_2 \left( 1 + \gamma_{\mathrm{c},u}(m) \right)
\end{equation}
The instantaneous rate for decoding the private stream at user $u$, where $0 \le u \le K_1-1$ is expressed as follows,
\begin{equation}
    R_{\mathrm{p},u} = \log_2 \left( 1 + \gamma_{\mathrm{p},u}(u) \right)
\end{equation}
The instantaneous rate at OTFS user $u, 0 \le u\le K_2-1$ is expressed as follows,
\begin{equation}
    R_{u} = \log_2 \left( 1 + \gamma_u \right).
\end{equation}
If the common stream is jointly encoded and then spread over all subcarriers, in order to ensure a successful decoding of the common stream at all users, we provide the following condition on the allocated rate for the common stream $R_\mathrm{c}$,
\begin{equation}\label{cond2}
    R_{\mathrm{c},u} \ge R_{\mathrm{c}} = \sum_{i = 0}^{K_1-1} C_i   , \forall \; 0 \le u \le K_1-1.
\end{equation}

Therefore, the achievable rate of OFDM user $u, 0 \le u\le K_1-1$ is $R_u = C_u +  R_{\mathrm{p},u}$.
If the Tx allocates the same rate for all portions of the common stream, meaning that $C_u = C, \; \forall \; 0 \le u \le K_1-1$, then, $C = {R_\mathrm{c}}/{K_1}$.

Herein, we assume that the necessary condition for a successful decoding of the common stream is $R_{\mathrm{c},u} \ge R_\mathrm{c}$. 
The outage probability of OFDM users implementing RSMA depends on the successful decoding of both the common stream and private stream at each user. We can consider the communication link of user $0 \le u \le K_1-1$ to be in outage in two cases:
\begin{itemize}
    \item If the common stream is not successfully decoded at user $u$, i.e., $R_{\mathrm{c},u} < R_\mathrm{c} = \sum_{u = 0}^{K_1-1} C_u$.
    \item If the common stream is successfully decoded at user $u$ but the overall rate of OFDM user $u$ is in outage after decoding the private stream, i.e, $C_u + R_{\mathrm{p},u} < R_0$.
\end{itemize}
These two outage events are statistically coupled because both the decoding of the common stream and private stream depend on the same channel realization on the subcarrier of the considered user. Moreover, the imperfect channel estimation translates into a residual interference during the SIC stage, leading to an inherent system sensitivity to estimation accuracy.

\section{Simulation Results}

\begin{figure}[!t]
	\centering 
	\captionsetup{justification = centering,margin = 1cm}
	\includegraphics[width = 3in]{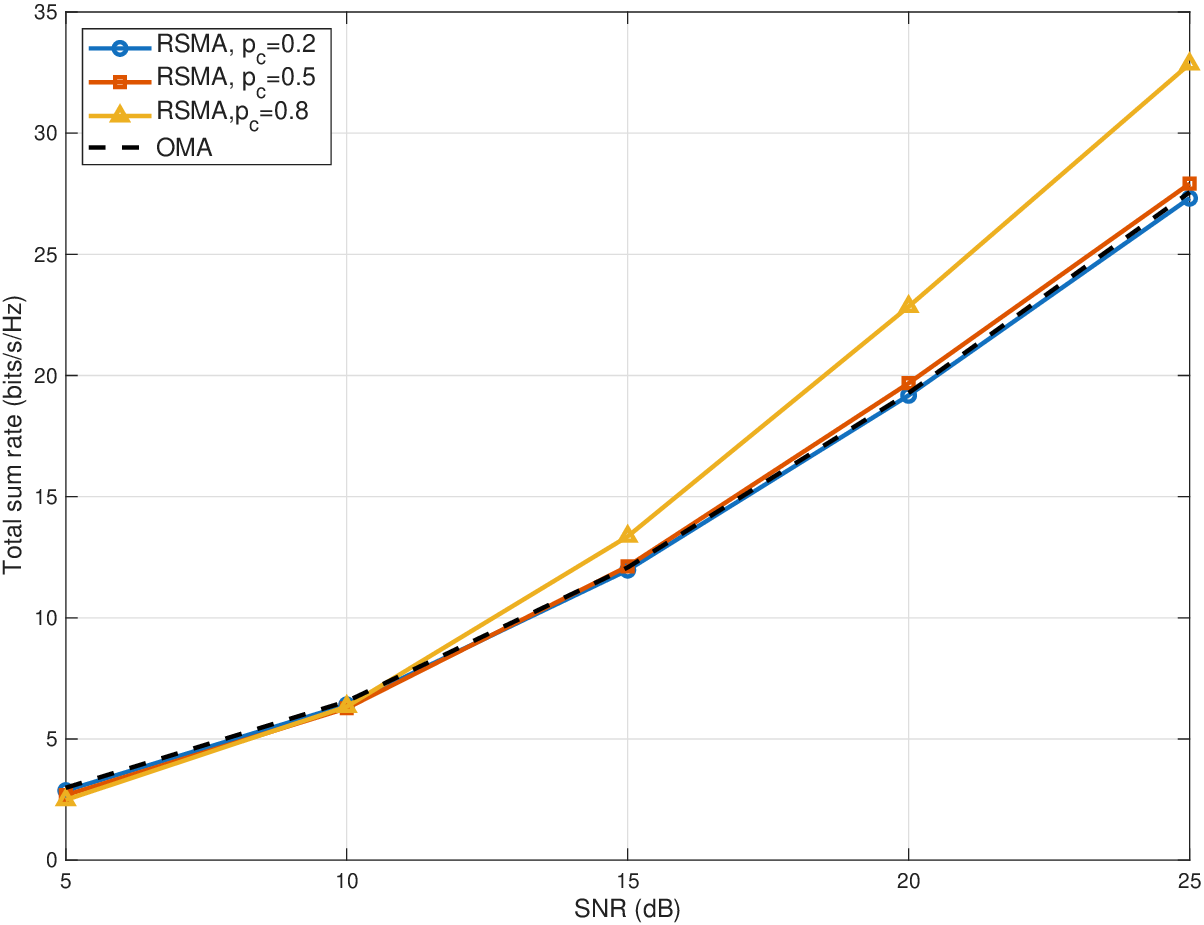}
	\caption{Total sum-rate vs SNR for different common stream power allocation factors.}
	\label{fig1}
\end{figure}
\begin{figure}[!t]
	\centering
	\captionsetup{justification=centering,margin=1cm}
	
	\subfloat[OFDM users sum-rate vs SNR.\label{fig2}]
	{
		\includegraphics[width=3in]{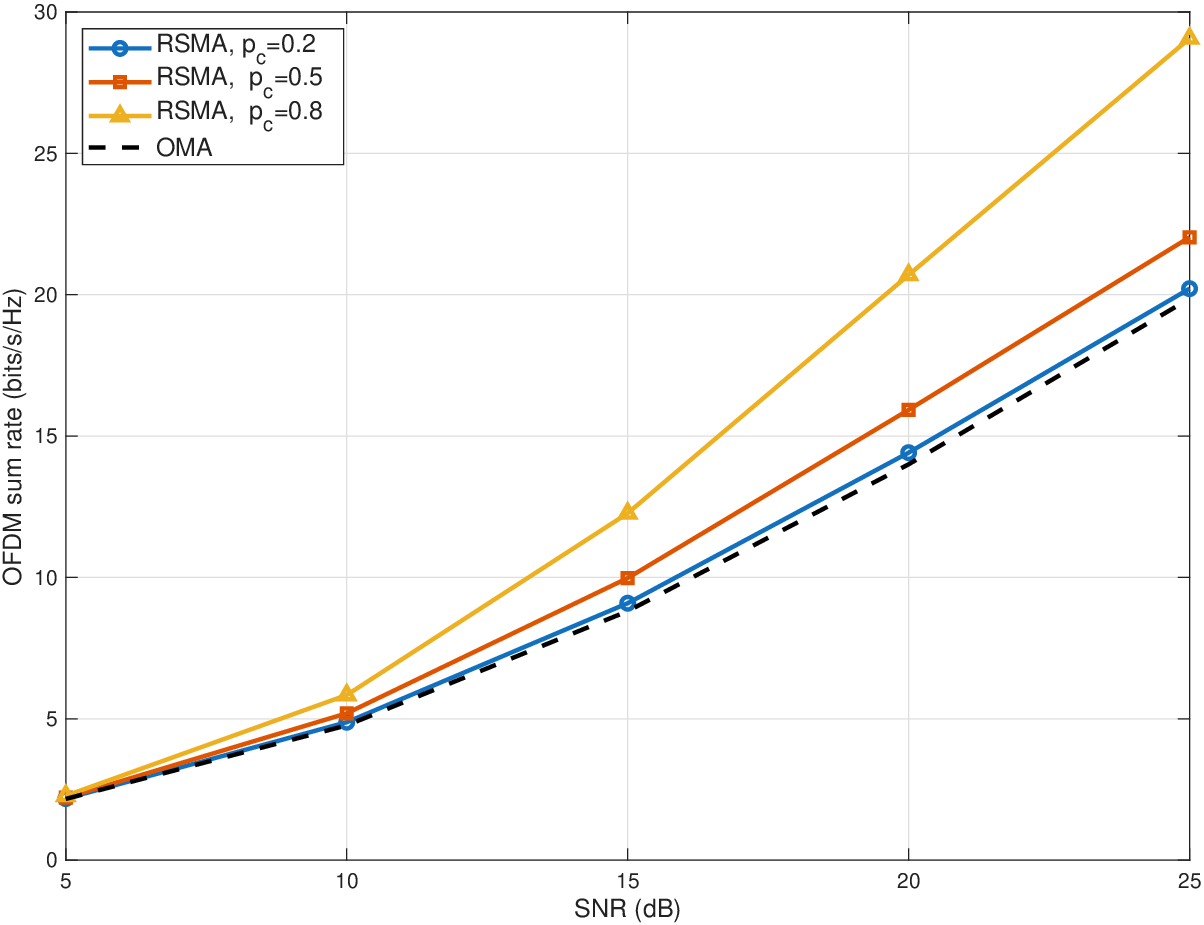}
	}
	
	\vspace{0.2cm}
	
	\subfloat[OTFS users sum-rate vs SNR.\label{fig3}]
	{
		\includegraphics[width=3in]{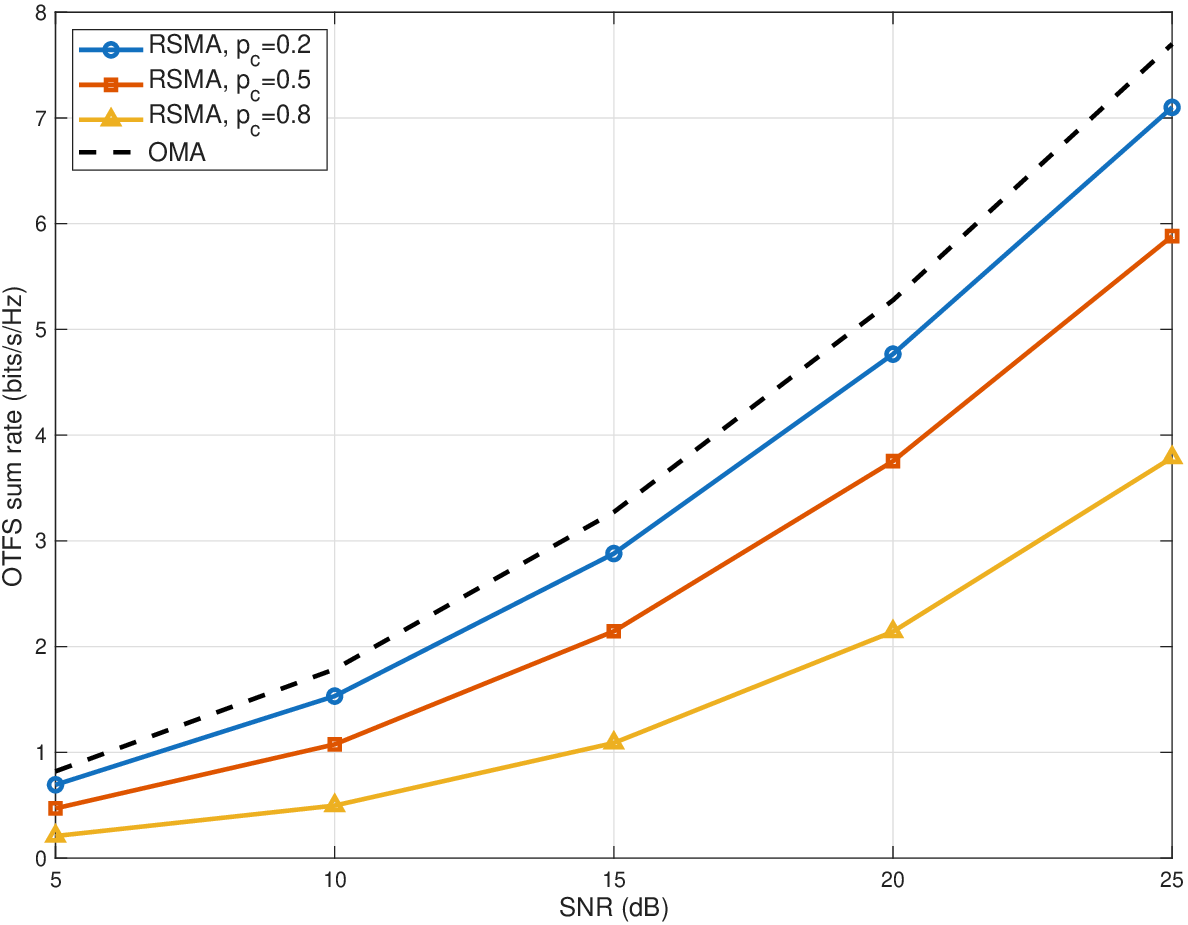}
	}
	
	\caption{Sum-rate performance of OFDM and OTFS users versus SNR for different common stream power allocation factors.}
	\label{fig:sumrate_snr}
\end{figure}
In this section, we evaluate the performance of the proposed RSMA-assisted OFDM-OTFS framework. The parameters used in the simulations are the following: $K_1=4,K_2=2,M=24,N=16,L=10, R_0 = 0.5 \; \text{b/s/Hz}$. The OFDM users experience mobility levels ranging from $40$ km/h to $120$ km/h. The OTFS users experience mobility levels ranging from $120$ km/h to $300$ km/h.
The results are obtained through $10^5$ Monte Carlo iterations over doubly selective channels with $P=4$ propagation paths for all users. The performance of the proposed RSMA framework is compared against an orthogonal multiple access (OMA) benchmark in the frequency domain. All parameters considered in the OMA benchmark are similar to the proposed RSMA setup including power and frequency allocations and ICI suffered by OFDM users.  In the following examples, the worst user indicates the user with the minimum average performance.

First, we evaluate the performance of the system in terms of total sum-rate, considering both OFDM and OTFS users. To this end, we plot in Fig. \ref{fig1} the total rate versus the SNR for different total common stream allocated power $p_\mathrm{c}$ and we compare it to the OMA benchmark. At low SNR values, all schemes exhibit comparable performance because the common stream contribution is limited at noise-limited regimes. However, at higher SNRs, the proposed RSMA system outperforms the OMA benchmark for common stream power share equal to 0.8. This behavior is justified by the ability of the RSMA architecture to improve the robustness of OFDM users which suffer from ICI. On the other hand, lower common stream power shares do not provide performance gain compared to OMA due to the reduced contribution of the common stream.
 Overall, the results demonstrate that the proposed framework becomes increasingly advantageous in high-SNR and mobility-sensitive scenarios when sufficient power is allocated to the common stream.

Second, we evaluate the sum-rate of each group of users separately. 
Fig. \ref{fig2} shows the OFDM sum-rate versus SNR for different common-stream power allocations. We observe that increasing the common-stream power allocation significantly improves the OFDM sum-rate performance since they directly benefit from the additional information transmitted through the RSMA layer over a wider range of frequencies. In fact, the common stream improves the robustness of OFDM users against mobility-induced ICI. As a result, the proposed framework increasingly outperforms OMA as the power factor $p_\mathrm{c}$ increases. For lower $p_\mathrm{c}$, the common stream contribution is limited.  
Fig. \ref{fig3} depicts the OTFS sum-rate versus SNR for different common-stream power allocations. Unlike the OFDM users, the OTFS users do not participate in the transmission of the common stream and only decode their private streams. Therefore,  their performance degrades when the common stream power allocation increases. Moreover, OMA slightly outperforms the proposed framework for OTFS users since the entire transmit power in OMA is dedicated to private messages and also due to the absence of SIC errors. These results highlight the trade-off introduced by the proposed RSMA-assisted architecture between improving OFDM robustness and preserving OTFS throughput performance.

\begin{figure}[!t]
	\centering
	\captionsetup{justification=centering,margin=1cm}
	
	\subfloat[Worst OFDM user rate vs common stream power for different SIC error variances.\label{fig4}]
	{
		\includegraphics[width=3in]{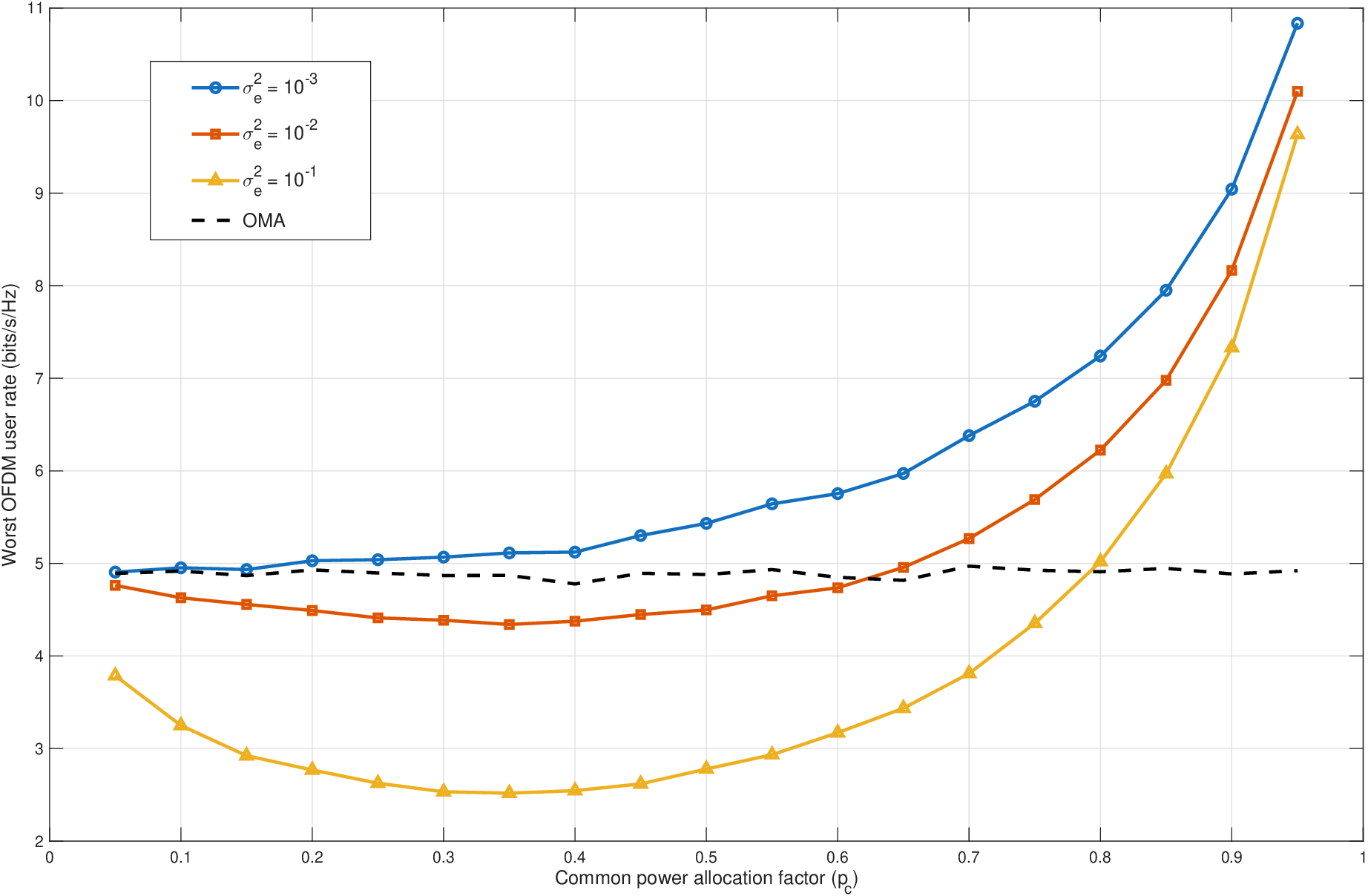}
	}
	
	\vspace{0.2cm}
	
	\subfloat[Worst OTFS user rate vs common stream power for different SIC error variances.\label{fig5}]
	{
		\includegraphics[width=3in]{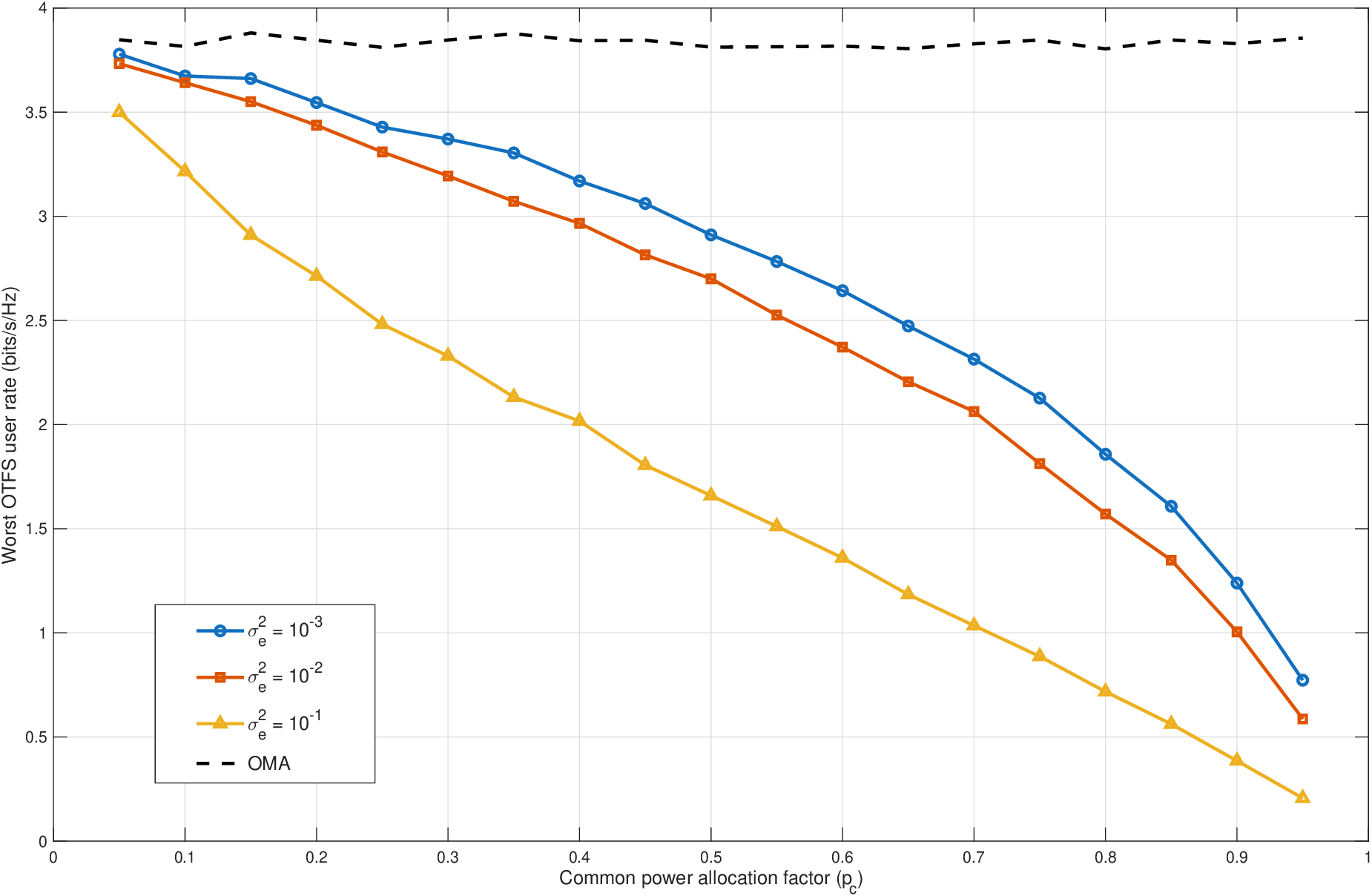}
	}
	
	\caption{Worst-user performance versus common stream power under different SIC error variances.}
	\label{fig:worstuser_pc}
\end{figure}

Third, we evaluate the impact of the common stream allocation factor and the SIC error variance on the performance of the proposed framework. Thus, in Fig. \ref{fig4}, we plot the worst OFDM user rate versus the total common-stream power allocation for different channel estimation error variances. For small channel estimation errors, increasing $p_\mathrm{c}$ improves the rate of the OFDM user and the RSMA system outperforms the OMA benchmark for all values of $p_\mathrm{c}$.
For more severe estimation errors, increasing the common stream power share degrades the performance of the OFDM user because of the significant residual interference induced by the imperfect SIC. This interference eventually becomes dominant. Nevertheless, at sufficiently large $p_\mathrm{c}$ values, the common stream compensates for the estimation errors and provides a significant performance gain over the OMA benchmark.
On the other hand, in Fig. \ref{fig5}, we plot the worst OTFS user rate versus the common stream power allocation factor for different channel estimation error variances. The OTFS user decodes and cancels the common stream without receiving a common message of its own. Thus, its performance degrades with the increase of $p_\mathrm{c}$.
 This degradation becomes more severe with increased channel estimation errors since the OTFS equalization becomes less effective under inaccurate channel knowledge and more residual interference. In contrast, OMA maintains a better performance because all the available transmit power remains dedicated to private transmissions. These results emphasize  the dependence of the proposed RSMA framework on the common stream power share and channel estimation quality.

\begin{figure}[!t]
	\centering
	\captionsetup{justification=centering,margin=1cm}
	
	\subfloat[OFDM user outage probability.\label{fig6}]
	{
		\includegraphics[width=3in]{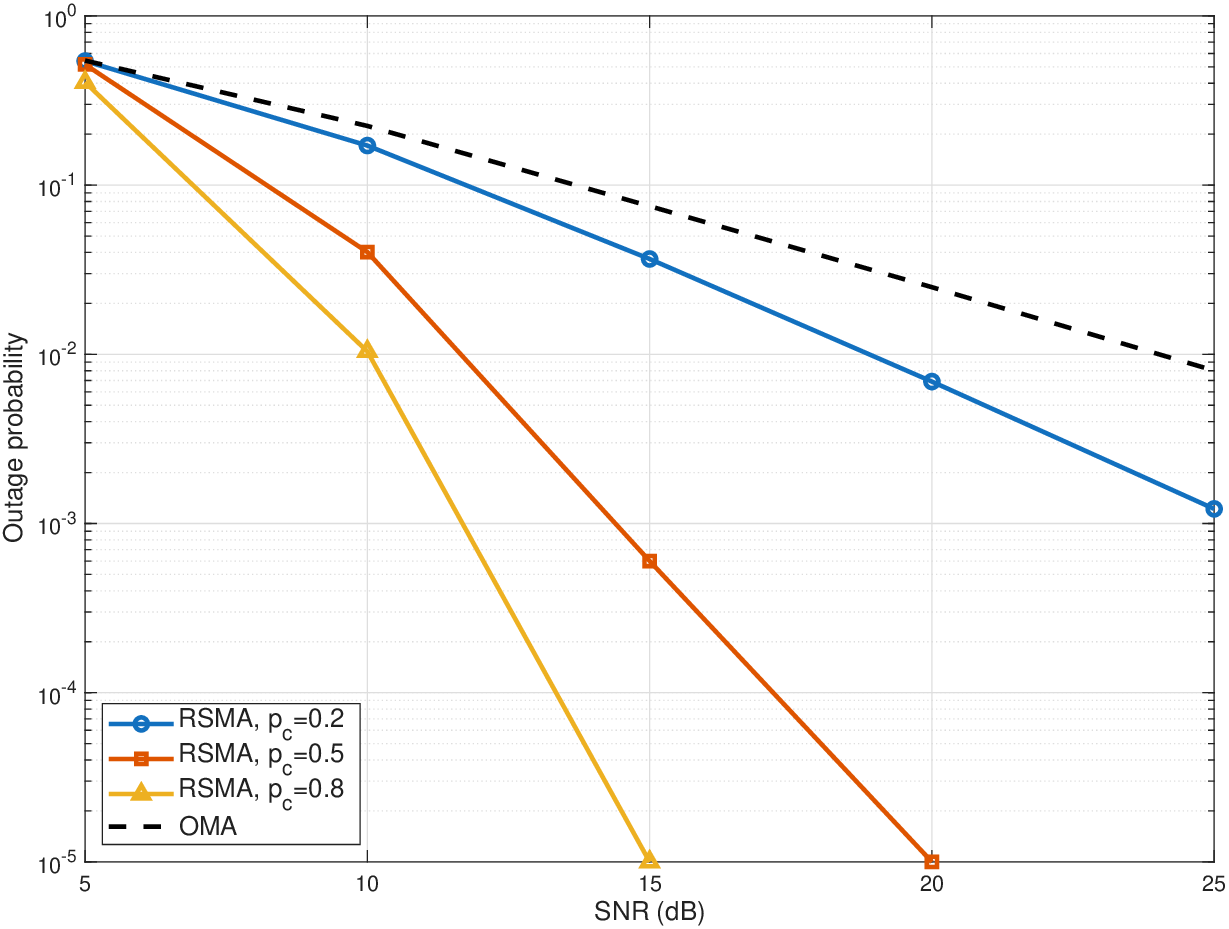}
	}
	
	\vspace{0.2cm}
	
	\subfloat[OTFS user outage probability.\label{fig7}]
	{
		\includegraphics[width=3in]{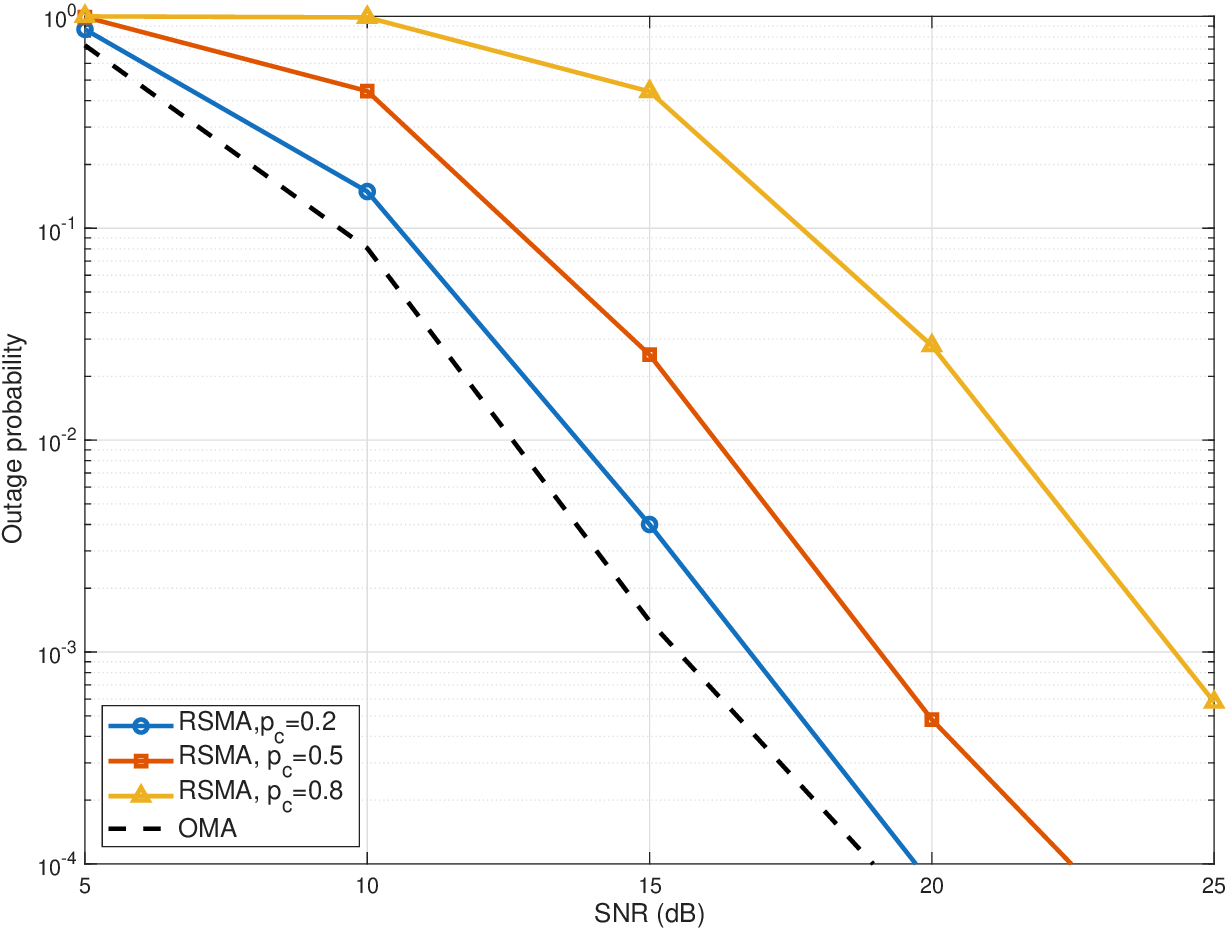}
	}
	
	\caption{Outage probability performance of OFDM and OTFS users for different common stream power allocation factors.}
	\label{fig:outage_users}
\end{figure}

 Last, we evaluate the performance of the proposed RSMA system in terms of outage probability. Therefore, we plot in Fig. \ref{fig6} the outage probability of the worst OFDM user versus the SNR for different common stream power allocation factors and we compare it to OMA. The OFDM user benefits from the increasing common stream share $p_\mathrm{c}$ and significantly outperforms its orthogonal counterpart. This gain is caused by the significant degradation of the performance of the OMA system due to ICI, while the RSMA system benefits from the common layer. 
  In particular, when $p_\mathrm{c} = 0.8$, the system achieves the best outage performance for the OFDM user and clearly outperforms OMA across most SNR values and for all values of estimation errors. We also evaluate the outage probability of the worst OTFS user in Fig. \ref{fig7}. Since the OTFS user benefits only from its private message, increasing $p_\mathrm{c}$ degrades its outage performance due to the reduced power share available for the OTFS user. Thus, the outage probability increases significantly as $p_\mathrm{c}$ increases from 0.2 to 0.8. This performance degradation is more noticeable at higher SNR values, where the system becomes increasingly power-allocation limited rather than noise-limited. In contrast, OMA achieves the lowest outage probability because all the transmit power is dedicated to the private message of the OTFS user. These results further confirm the trade-off introduced by the proposed RSMA-assisted framework between improving OFDM robustness and preserving OTFS reliability.

\section{Conclusion}
In this paper, we proposed a novel RSMA-assisted transmission framework to enable the coexistence of OFDM and OTFS waveforms in mixed mobility communication networks. Numerical results demonstrated that the proposed framework provides significant improvements in OFDM user rate and outage performance when sufficient power is allocated to the common stream. In particular, increasing the common stream power allocation substantially enhanced OFDM reliability and throughput especially at high SNR regimes. The results also revealed an inherent trade-off between OFDM and OTFS performance, since allocating excessive power to the common stream reduces the power available for OTFS transmissions and may degrade OTFS throughput and outage performance. Furthermore, the impact of channel estimation errors on both OFDM and OTFS users showed that the proposed framework remains beneficial under different estimation inaccuracies with suitable power distribution.
\bibliographystyle{IEEEtran}
\bibliography{IEEEabrv,main}
\end{document}